# Computational Microwave Imaging Using 3D Printed Conductive Polymer Frequency-Diverse Metasurface Antennas

Okan Yurduseven[1,*], Patrick Flowers[2], Shengrong Ye[2], Daniel L. Marks[1], Jonah N. Gollub[1], Thomas Fromenteze[3], Benjamin J. Wiley[2], and David R. Smith[1]

[1]Center for Metamaterials and Integrated Plasmonics and Department of Electrical and Computer Engineering, Duke University, Durham, North Carolina 27708, United States
[2]Department of Chemistry, Duke University, Durham, North Carolina 27708, United States
[3]Xlim Research Institute, University of Limoges, 87060 Limoges, France
[*]okanyurduseven@ieee.org

**Abstract:** A frequency-diverse computational imaging system synthesized using three-dimensional (3D) printed frequency-diverse metasurface antennas is demonstrated. The 3D fabrication of the antennas is achieved using a combination of PolyLactic Acid (PLA) polymer material and conductive polymer material (Electrifi), circumventing the requirement for expensive and time-consuming conventional fabrication techniques, such as machine milling, photolithography and laser-etching. Using the 3D printed frequency-diverse metasurface antennas, a composite aperture is designed and simulated for imaging in the K-band frequency regime (17.5-26.5 GHz). The frequency-diverse system is capable of imaging by means of a simple frequency-sweep in an-all electronic manner, avoiding mechanical scanning and active circuit components. Using the synthesized system, microwave imaging of objects is achieved at the diffraction limit. It is also demonstrated that the conductivity of the Electrifi polymer material significantly affects the performance of the 3D printed antennas and therefore is a critical factor governing the fidelity of the reconstructed images.

## 1. Introduction

Imaging at microwave and millimeter-wave frequency regimes has recently received considerable attention in the literature. Radiation in these frequency bands is non-ionizing and can penetrate through most optically opaque materials, and is thus ideally suited for a variety of emerging imaging applications, including through-wall imaging [1, 2], non-destructive testing [3, 4], biomedical imaging [5, 6], and security-screening [7-10].

Investigating the literature, conventional imaging modalities used in these applications can be understood as versions of synthetic aperture radar (SAR) [1, 3-10] and phased array (or electronic beam scanning) [2, 11-13] systems. Using these techniques, high fidelity imaging has been demonstrated. Conventionally, these techniques interrogate the scene to be imaged at the Nyquist limit ($\lambda_0/2$), where $\lambda_0$ is the free-space wavelength. The fields produced in such systems are essentially orthogonal, achieved by means of mechanical scanning in SAR or electronic beam forming in phased array systems.

While capable of producing high-fidelity images, both phased arrays and SAR systems exhibit significant limitations. In SAR, for example, an antenna (or an array of antennas) is mechanically translated to synthesize a composite aperture, limiting the data acquisition speed. This can become a considerable challenge for applications where imaging is required to be performed over a large field-of-view (FOV). Phased arrays, on the other hand, have the potential to address this challenge by offering all-electronic operation. Conventionally, in phased arrays, a composite aperture is synthesized using an array of antennas with the radiated fields steered through the use of phase shifters located behind each of the array antennas. For an aperture synthesized at the Nyquist limit, the number of antennas required for imaging can be significant, especially for applications where a large FOV is required. Moreover, to have the full phase control of the individual antennas within the synthesized aperture, phase shifting and switching circuits are needed, resulting in power amplifiers and other active components being used to compensate for the insertion loss of these circuits. Thus, phased array systems can be expensive, cumbersome and exhibit complex hardware architecture.

Recently, the concept of frequency-diversity leveraging computational imaging has been shown to be a promising alternative to address these challenges [14-28]. Computational imaging techniques enable the system hardware architecture to be simplified by moving the burden from the hardware layer to the image processing (software) layer [29-33]. With the developing computing power of modern computers and the implementation of general purpose graphics processing units (GPGPU) for imaging [25], modern computers have the capability to solve problems of increasing complexity more than ever.

Frequency-diversity is an all-electronic technique, enabling the imaging to be performed by means of a frequency sweep, with no mechanically moving parts or active circuit components required. In this technique, frequency-diverse antennas are used to synthesize a composite aperture, interrogating the scene. The frequency-diverse antennas radiate quasi-orthogonal field patterns that vary strongly as a function of the driving frequency. Therefore, as the imaging frequency is swept over a given frequency band, scene information is encoded onto a set of measurements taken at a set of sampled frequency points. Using these measurements, the scene is reconstructed leveraging computational imaging algorithms.

Antenna choice is an important aspect for the design of a frequency-diverse imaging system. A critical factor governing the performance of an antenna for computational imaging is its quality (Q-) factor. The Q-factor of an antenna governs the orthogonality of the radiated fields produced by the antenna as a function of frequency. For a frequency-diverse antenna, it is desirable to maximize the Q-factor, minimizing the overlap between the radiated field patterns sampling the scene at adjacent frequencies. The overall coverage in the spatial frequency domain (k-space) is determined by the convolution of the radiated fields from a transmit and receive pair of antennas; minimizing correlation between the measurements amounts to ensuring that the convolution of the fields covers as large a region of k-space as possible with minimal redundancy [14, 15].

Usually frequency-diverse antennas are fabricated using machining, photolithography and laser printing. Although these manufacturing tools are highly accurate and reliable, they can be expensive and time-consuming. For example, fabricating the frequency-diverse antennas presented in [26] requires the machining of the metal structure by removing a large amount of material from a large piece of metal. It also results in a heavy and bulky antenna. The alternative printed antennas reported in [15, 16, 22, 23] require the use of high-precision printed circuit board (PCB) and laser etching printers, which are expensive. Moreover, whereas such printers are suitable to fabricate planar structures, they are not convenient for 3D designs.

Leveraging the concept of 3D printing, structures – even with complex shapes – can be realized by means of additive manufacturing, using the fused deposition method (FDM) [34, 35] or the polymer jetting (polyjet) method [36, 37]. A significant advantage of the 3D printing technique is that fabrication of the antennas can be achieved rapidly, without the need for conventional machining approaches. Despite these advantages, 3D printing technology for RF applications brings its own challenges. 3D printers historically have made use of nonconducting polymer materials; whereas for RF applications, conductive structures are required. Thus, 3D printed components requiring conducting regions have typically used metallization via plating methods—an approach that has considerable constraints in terms of the types of elements that can be rendered conducting.

In this paper, we demonstrate an alternative 3D printing approach using conductive polymer as the model material for fabrication, circumventing the requirement to adopt additional metallization techniques for RF applications. The conductive material, which we refer to as *Electrifi* [38], is a metal-polymer composite that consists of a biodegradable polyester and copper. It has a resistivity of $6*10^{-5}$ Ω/m (or a conductivity of $\sigma=1.67*10^4$ S/m), and is compatible with the most commercial desktop FDM 3D printers. Using the conductive polymer material, we manufacture frequency-diverse metasurface antennas and demonstrate the application of such antennas for computational imaging applications. Using an in-house developed simulation code, termed the Virtualizer [18], a composite aperture is synthesized by employing an array of the 3D printed frequency-diverse metasurface antennas for imaging over the K-band frequency regime (17.5-26.5 GHz).

The outline of the paper is as follows. In section 2, the concept of frequency-diverse imaging is explained. The inverse problem, imaging equation and computational techniques for image reconstruction are introduced. Section 3 discusses the 3D fabrication of the frequency-diverse antennas used to synthesize the composite aperture and the K-band imaging results for a number of resolution-analysis objects; a point-scatter array and a resolution target. The effect of the material conductivity on antenna performance and quality of the reconstructed images is also demonstrated. Finally, section 4 provides the concluding remarks.

## 2. Frequency-Diverse Imaging

The concept of frequency-diverse imaging relies on encoding scene information onto a set of frequency points. Operating over a certain frequency band, as the imaging frequency is swept, the fields radiated by the antennas sample the scene, which is reconstructed using computational imaging algorithms.

Reconstructing the scene from a set of measurements is an inverse problem, requiring a model to be established between the fields radiated by the antennas, the scene to be reconstructed and the measured return signal collected from the scene. In this work, we refer to this process as the *forward model*. The diffraction limit associated with the finite aperture, as well as the finite frequency bandwidth of operation imply that the scene can be discretized into a set of voxels, each of which is connected to the set of measurements by the matrix equation [17]:

$$\mathbf{g}_{Mx1} = \mathbf{H}_{MxN}\mathbf{f}_{Nx1} + \mathbf{n} \tag{1}$$

In (1), $\mathbf{g}$ is the measured signal, $\mathbf{H}$ is the measurement (or sensing) matrix, $\mathbf{n}$ is the system noise and $\mathbf{f}$ is the scene reflectivity (or contrast) vector while $M$ and $N$ denote the number of total measurement modes and the number of three-dimensional (3D) voxels into which the scene is discretized. The measurement matrix $\mathbf{H}$ is the product of the fields radiated by the transmit, $\mathbf{E}_{Tx}$, and receive, $\mathbf{E}_{Rx}$, antennas respectively. To characterize the fields radiated by the antennas, we make use of a near-field scanning system, NSI 200V-3x3, as depicted in Fig. 1 [15, 21]. By scanning the fields over a plane near the aperture, we can determine the fields everywhere throughout the measurement volume by the aperture fields to the desired points. The near-field scanning approach to characterization obviates the need for full-wave simulations or for analytical models of the antenna properties.

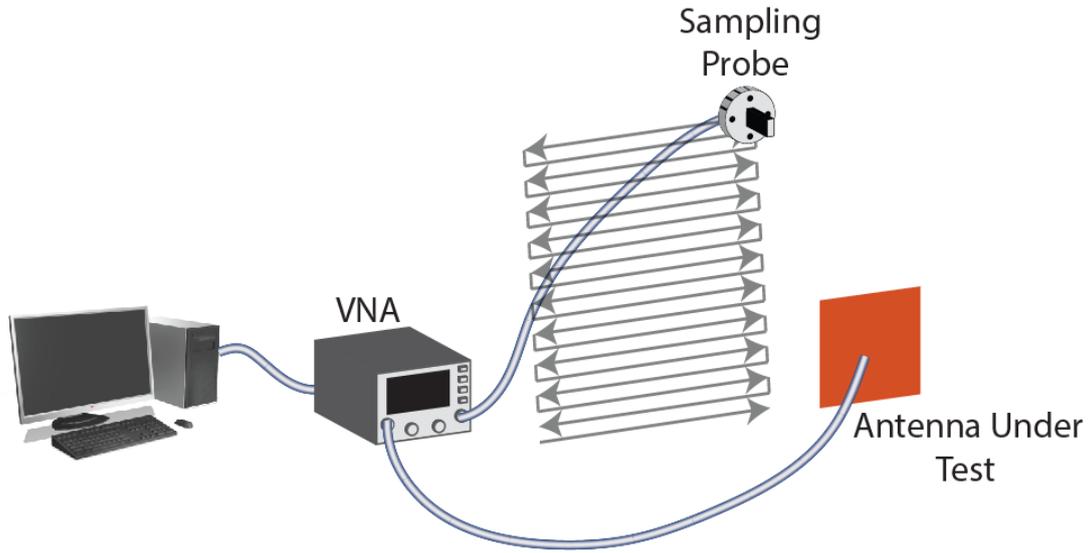

*Fig. 1. Characterization of a frequency-diverse antenna by means of near-field scanning*

Investigating the size of the measurement matrix, **H**, it is evident that the imaging problem can be over-determined (*M>N*) or under-determined (*M<N*). As the measurement matrix is not square, and therefore does not have an inverse, solving (1) for **f** does not have an exact solution. As a result, computational techniques can be used to reconstruct an estimate of the scene, **f**$_{est}$, from (1). A number of computational techniques have been reported in the literature, from single-shot reconstruction algorithms, such as pseudo-inverse and matched-filter (that require a single matrix multiplication), to more advanced iterative algorithms, such as least-squares, two-step iterative shrinkage/thresholding algorithm (TwIST) and TwIST with total variation regularization (TwIST+TV) [20, 39, 40]. Single-shot reconstruction techniques are not iterative and therefore are advantageous in that they are computationally inexpensive, suggesting that these techniques lead to fast reconstruction times. However, iterative reconstruction algorithms can reconstruct better quality image estimations in comparison to single-shot algorithms, thus introducing a tradeoff between reconstruction quality and the reconstruction time. As a result, for applications where imaging quality is more important than the reconstruction speed, iterative techniques can be employed while for applications requiring fast reconstruction, such as real-time imaging, single-shot reconstruction algorithms can be more desirable. In this work, we make use of the least-squares algorithm for image reconstruction.

## 3. Antenna Fabrication and Imaging Results

To obtain the greatest diversity of radiation patterns from a frequency-diverse antenna, a number of parameters must be optimized, such as the quality factor (Q-factor), radiation efficiency, and Fourier space (k-space) sampling [15]. The Q-factor plays the dominant role in determining the orthogonality of the fields radiated by the antennas at adjacent frequency points over the operating frequency band. Increasing the Q-factor is desirable in that it reduces the spatial overlap between radiation patterns, reducing the redundancy of the information collected from the scene as the frequency is swept. However, the Q-factor of a frequency-diverse antenna is inversely proportional to the radiation efficiency, which governs the signal-to-noise ratio (SNR) for imaging [21, 23]. As a result, there is a tradeoff between the Q-factor and the radiation efficiency of a frequency-diverse antenna, which needs to be tailored for the requirements of the desired application. Because the Q-factor is such an important parameter for imaging applications, the use of cavity-backed apertures can be desirable for frequency-diverse antennas. The larger the volume of the cavity, the larger the Q-factor and the larger the number of usable radiation patterns available. In practice, the total number of useful radiation patterns for imaging is limited by the space bandwidth product (SBP) or, equivalently, the diffraction limit associated with the aperture size.

In recent work [14, 15, 26], we have reported the concept of cavity-backed, Mills-Cross metasurface antenna and demonstrated that these antennas have optimum characteristics for frequency-diverse imaging. A depiction of the Mills-Cross structure is shown in Fig. 2.

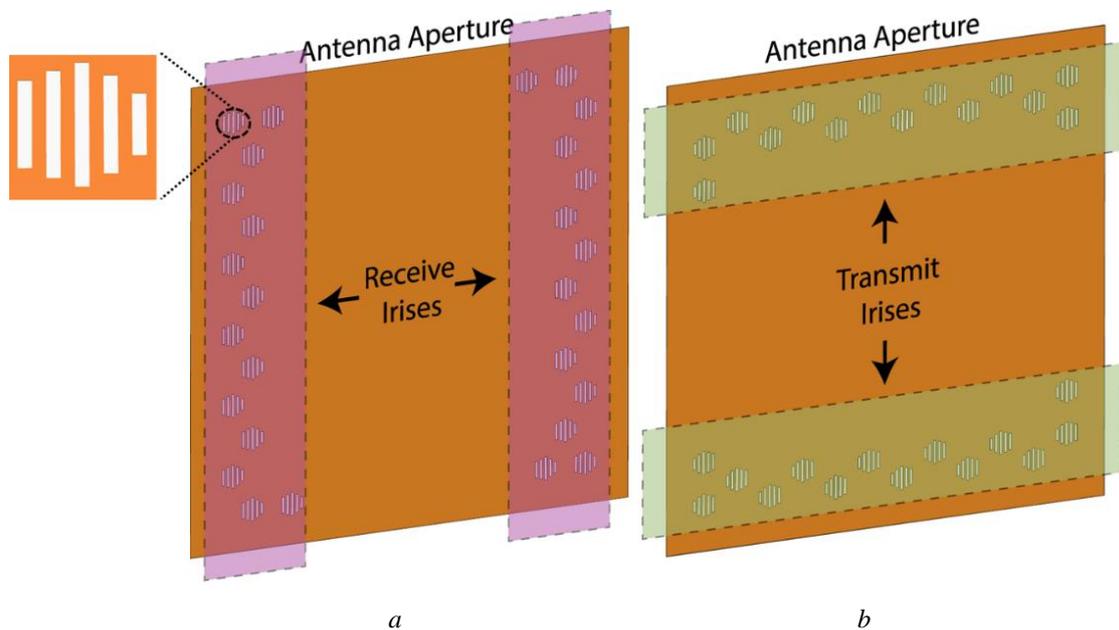

*Fig. 2. The Mills-Cross configuration. A close look-up at a single iris (or unit cell) is also shown. The antennas can be planar (2D) or volumetric (3D)*
*a* Receive antenna with the irises oriented along the vertical axis
*b* Transmit antenna with the irises are oriented along the horizontal axis

As can be seen in Fig. 2, in the Mills-Cross configuration, the radiating irises (or unit cells) on the receive and transmit antennas are perpendicular to each other and, if overlapped, form a Mills-Cross pattern for a given transmit and receive antenna pair. We have demonstrated both planar printed circuit board (PCB) as well as air-filled cavity versions of the Mills-Cross antennas [15, 26], with the latter having significantly larger Q-factor while achieving moderate radiation efficiencies. The radiating irises are sub-wavelength in size and can be circular slots for polarimetric imaging [15, 21-23] or rectangular slots for single polarization imaging [24, 26, 27]. The iris structure depicted in Fig. 2, which is also adopted in this work, consists of multiple sub-wavelength slots of varying lengths (on the order of $\lambda/4$ - $\lambda/2$ over the K-band) in order to flatten the radiation response of the antennas over the K-band frequency range [26].

We make use of the air-filled Mills-Cross cavity antenna design here to illustrate the unique 3D printing approach. The Mills-Cross antennas developed in [26] were machined from an aluminum block, in which most of the material was removed by a computer-controlled milling machine. As an alternative, in this work, we fabricate the frequency-diverse antennas using a 3D printer, leveraging the FDM method, resulting in the antennas being fabricated on a layer-by-layer basis. The 3D printer is custom-made, which was developed based on an open-source D-Bot design [41]. The custom-made printer has a number of important modifications in comparison to the original open-source version. The first modification is the conversion to dual direct-drive extrusion to prevent the possibility of the Electrifi material jamming in a Bowden tube. The second modification is the incorporation of AutoLift retractable all-metal hotends [42], which helps to eliminate cross-contamination of print materials.

The 3D printer supports dual-mode printing, enabling two different model materials to be used for the fabrication. As a result, the outer part of the cavity was fabricated using dielectric PLA model material while the inner walls of the cavity were covered using the Electrifi conductive polymer material, as shown in Figs. 3*a* and 3*b*. The overall wall thickness of the 3D printed cavity is 10 mm, ensuring that the fabricated prototype is rigid, while the wall thickness for the conductive part is 1 mm, significantly reducing the 3D printing cost of the antenna. The wall thickness for the conductive part was chosen to be larger than the skin depth of the Electrifi material determined by the conductivity, $\sigma=1.67*10^4$ S/m.

The PLA material was printed at 190 $^0$C at a speed of 30 mm/s while the Electrifi material was printed at 140 $^0$C at 15 mm/s speed. A 0.5 mm diameter nozzle was used for both materials and layer height was set at 0.2 mm. It should be noted that no heated bed was used for printing in order to maintain maximum conductivity of the Electrifi material. In order to prevent warpage of PLA, a number of stress relief structures, including chamfered edge and cylindrical voids near the corners of the antenna model, were included in the model. BuildTak [43] was used as the primary build platform, as it is compatible with PLA when no heated bed is used.

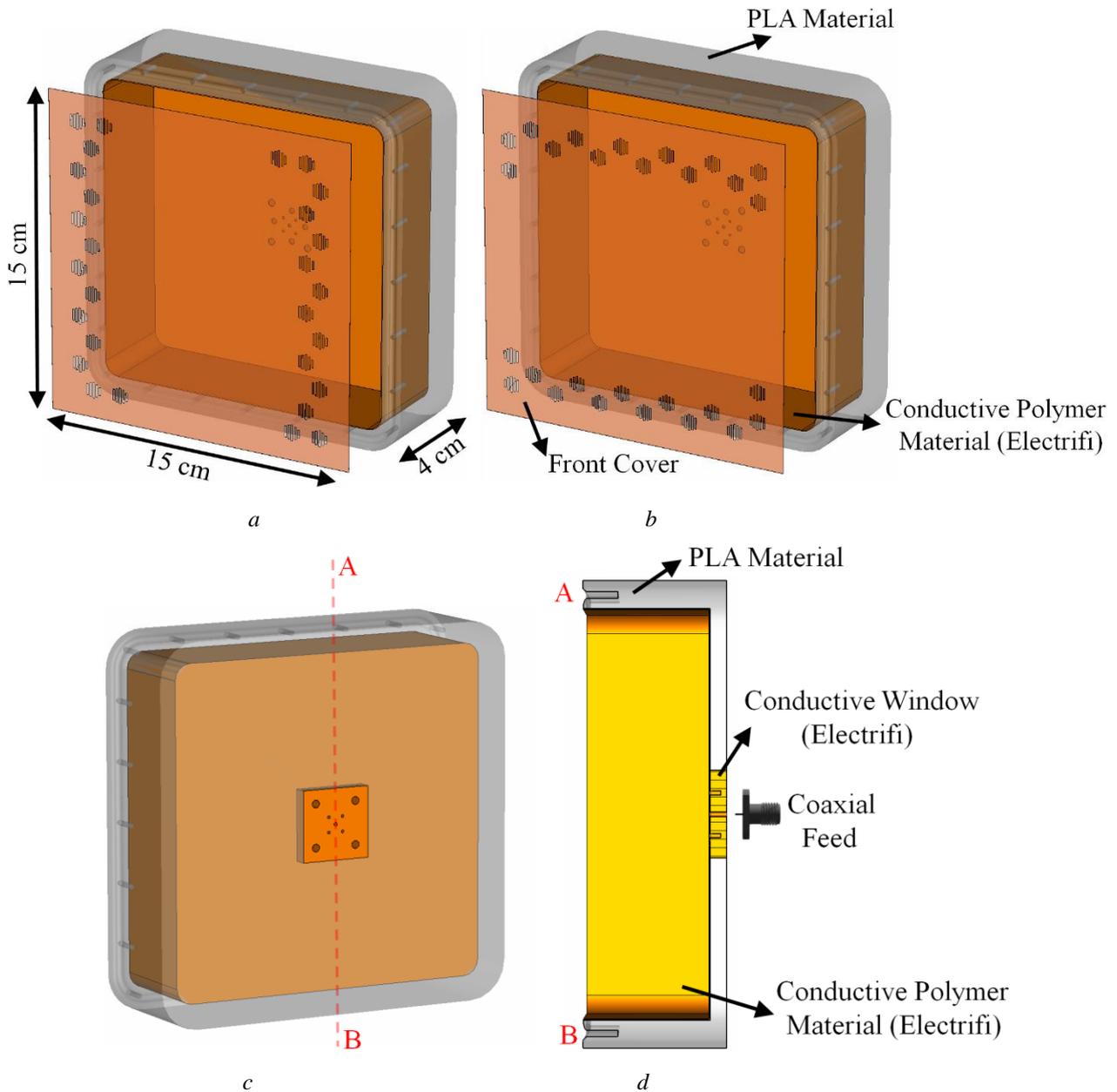

*Fig. 3.* *3D design of the air-filled Mills-Cross cavity metasurface antennas. Parts of the model printed with PLA and conductive polymer material (Electrifi) are highlighted*
*a* Front-view (receive antenna)
*b* Front-view (transmit antenna)
*c* Back-view
*d* Cross-section

As can be seen in Fig. 3*d*, the cavity is fed in the center using a coaxial feed. As depicted in Figs. 3*c* and 3*d*, in order to ensure that the wave is launched directly into the cavity and not into the dielectric PLA part, a conductive window was designed to guide the wave launched by the coaxial feed into the cavity. The fabricated 3D printed air-filled Mills-Cross antennas are shown in Fig. 4.

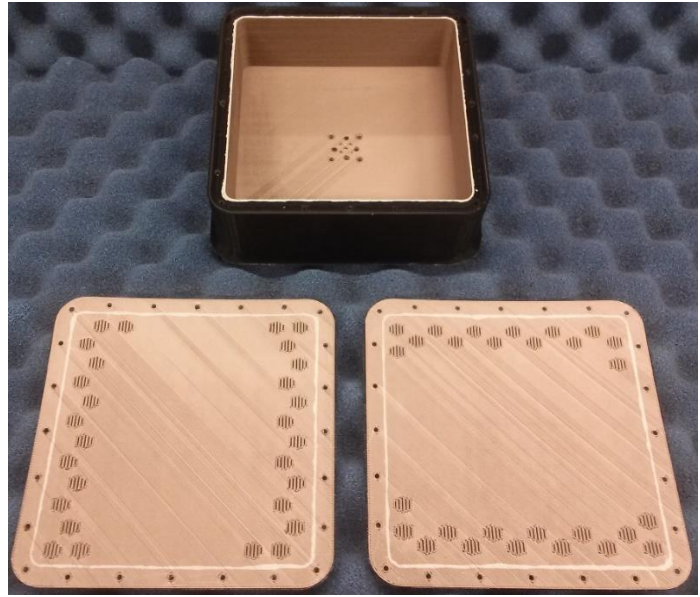

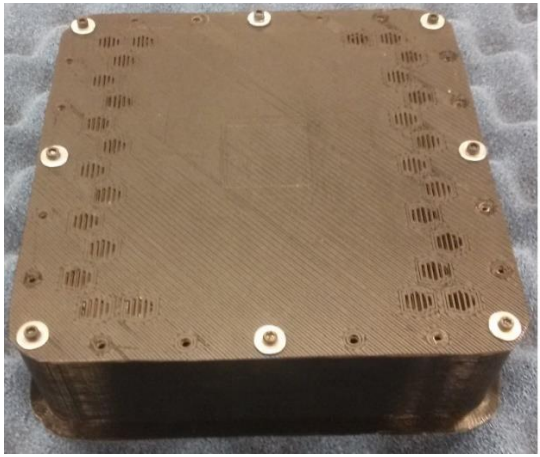
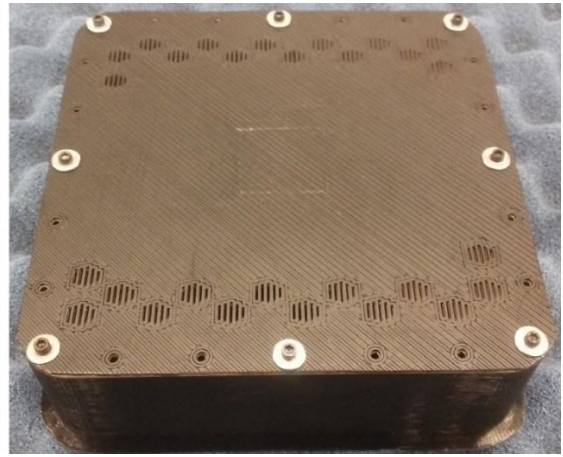

***Fig. 4.*** *Fabricated 3D printed Mills-Cross cavity antennas*
*a* Cavity base (top) with receive (bottom left) and transmit (bottom right) front covers
*b* Assembled receive cavity
*c* Assembled transmit cavity

Following the fabrication, the quality factor (Q-factor) of the antennas was analyzed. The Q-factor of a frequency-diverse antenna can be investigated by analyzing the impulse response of the antenna in the time domain [15]. Increasing the Q-factor results in a widened impulse response. Fig. 5*a* demonstrates the impulse response of the antennas measured using a vector network analyzer (VNA, Keysight N5222A) while the reflection coefficient pattern of the antennas across the K-band is shown in Fig. 5*b*. From the measured impulse-response pattern in Fig. 5*a*, the Q-factor of the 3D fabricated Mills-Cross antennas was calculated to be *Q*=300.

A key parameter in the 3D fabrication of the Mills-Cross cavity antennas is the conductivity of the Electrifi conductive polymer material. For an imaging system synthesized using these antennas, it is important that the effect of material conductivity on the antenna performance is investigated. In view of this, we performed the full-wave simulations of the antennas in CST Microwave Studio and analyzed the impulse response patterns of the antennas as a function of different conductivity values; a) reducing the conductivity of the Electrifi material by a factor of 10 ($\sigma=1.67*10^3$ S/m), b) using the actual conductivity value of the Electrifi material ($\sigma=1.67*10^4$ S/m), and c) increasing the conductivity of the Electrifi material by a factor of 10 ($\sigma=1.67*10^5$ S/m). The obtained impulse response patterns as a function of conductivity are also shown in Fig. 5*a*.

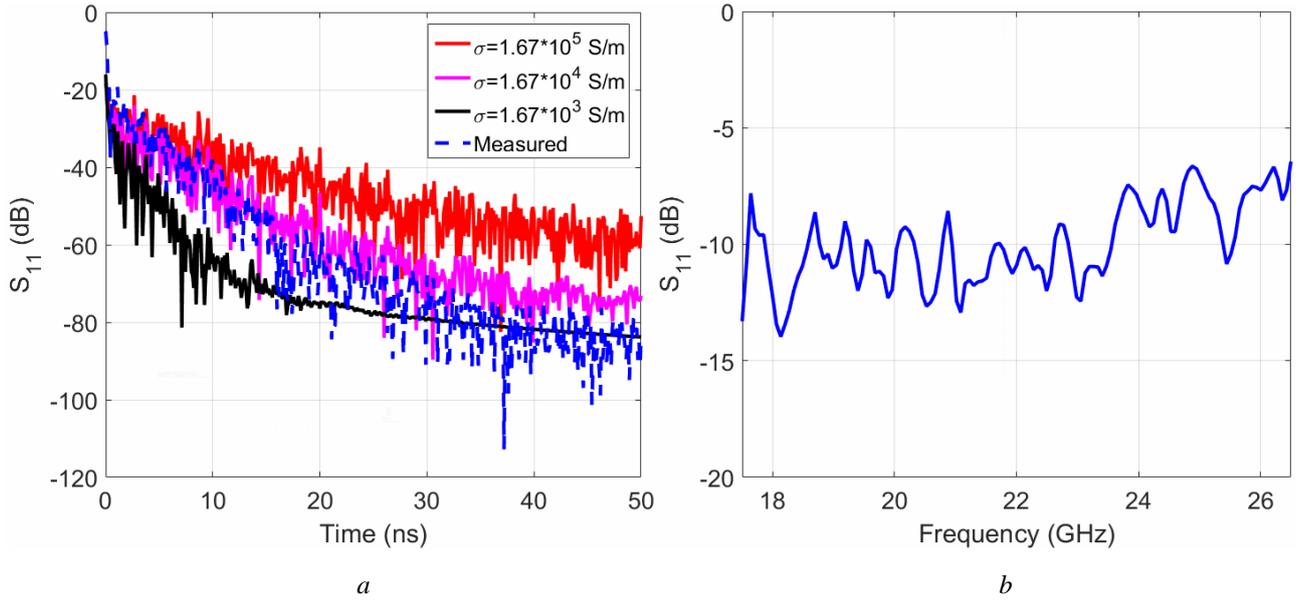

***Fig. 5.*** *$S_{11}$ response of the 3D printed Mills-Cross metasurface cavity*
*a* Simulated and measured time-domain patterns as a function of material conductivity (σ)
*b* Measured frequency-domain pattern

It can be seen in Fig. 5*a* that using the actual conductivity value of the Electrifi material ($\sigma=1.67*10^4$ S/m), good agreement is achieved between the simulated and measured impulse response patterns. Analyzing Fig. 5*a*, it is evident that the impulse response is narrower for $\sigma=1.67*10^3$ S/m and wider for $\sigma=1.67*10^5$ S/m, with the corresponding Q-factors were measured to be *Q*=150 and *Q*=800, respectively.

Determining the Q-factor of a frequency-diverse antenna enables the calculation of another important system parameter for imaging; the number of points used to sample the operating frequency band [14, 15], according to:

$$N_s = \frac{QB}{f_c} \qquad (2)$$

In (2), $B$ is the operational bandwidth while $f_c$ denotes the center imaging frequency. The K-band bandwidth is $B=9$ GHz with a center frequency of $f_c=22$ GHz. From (2), the optimum number of frequency sampling points, $N_s$, is calculated as 122. In this work, the K-band was slightly oversampled with the number of frequency points was chosen to be 201, resulting in $\Delta f=45$ MHz separation between the adjacent frequency points. Increasing the number of frequency sampling points beyond this limit would result in redundant information being collected from the scene and increase the size of the data set without any advantage. In order to demonstrate the variation of the fields radiated by the antennas, in Fig. 6, the measured field patterns are shown at three adjacent frequency points, centered at 22 GHz within the K-band.

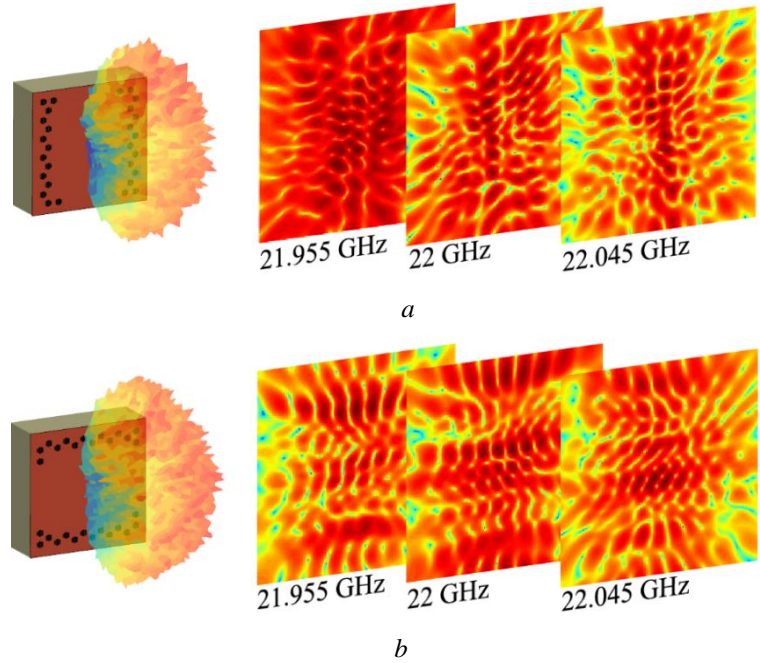

*Fig. 6. Electric field patterns radiated by the 3D printed antennas propagated to a distance of d=0.5 m over a 2 m x 2 m FOV. Field patterns are shown at adjacent frequency points, centered at 22 GHz. The rapid variation of the fields is evident*
*a* Receive antenna
*b* Transmit antenna

Following the calculation of the Q-factor and the number of frequency sampling points, the radiation efficiency of the antennas was investigated. The radiation efficiency was calculated by means of analyzing the radiated fields measured using the near-field scanning system [15, 21] and is reported to be $\eta=20\%$.

In order to use the 3D printed Mills-Cross antennas for imaging, a composite aperture needs to be synthesized. To this end, we use the Virtualizer [18], an in-house code developed in the Matlab programming environment. Using the Virtualizer, we can model composite frequency-diverse apertures of any desired size and configuration (monostatic, bistatic and multistatic), by means of either analytically modelling or importing the near-field scans of the fabricated antennas. Using the analytically calculated or measured fields in conjunction with Virtualized targets – collections of voxels each with assigned value of reflectivity – we can obtain accurate predictions of imaging performance. In view of this, we first import the experimental near-field scans of the 3D printed transmit and receive Mills-Cross antennas into the Virtualizer. At this stage the overall aperture consists of only two antennas. We then synthesize a larger multi-static aperture by populating an area with another pair of these antennas, resulting in a composite aperture depicted in Fig. 7*a*.

Using the Virtualizer, the near-field scans of the antennas are first back-propagated to the aperture plane of the antennas. The back-propagated fields are then modeled as an array of radiating magnetic dipoles, from which the electric field patterns can be calculated at any point in the scene using dyadic Green's functions [17]. The product of the fields from a given transmit antenna with those from a receive antenna form the measurement matrix, **H**, relating the scene reflectivity values to the measurements, as in (1). The total number of measurement modes supported by the imaging system depicted in Fig. 7*a* can be given as *M = number of transmit antennas* x *number of receive antennas* x *number of frequency sampling points*. The frequency-diverse antennas operate over the K-band (17.5-26.5 GHz), sampled at 201 frequency points, bringing the total number of measurement modes to *M*=804.

The resolution limit is one of the key metrics in defining the performance of an imaging system. To confirm that the imaging is done at the diffraction limit of the synthesized aperture, we analyze the point-spread-function (PSF) of the aperture by imaging an array of point sources shown in Fig. 7*a*. The scene is discretized into 3D voxels, with the dimensions of the voxels selected in accordance with the theoretical resolution limit of the synthesized composite aperture in range (x-axis), $\delta_r$, and cross-range (y-z plane), $\delta_{cr}$, calculated using SAR resolution equations as follows [20, 44, 45].

$$\delta_r = \frac{c}{2B} \qquad (3)$$

$$\delta_{cr} = \frac{\lambda_{min} R}{2D} \qquad (4)$$

In (3), *c* denotes the speed of light and *B* is the operating bandwidth (*B*=9 GHz). In (4), $\lambda_{min}$ is the free-space wavelength at 26.5 GHz and *R* is the approximate distance of the target from the aperture (*R*=50 cm), while *D* denotes the size of the overall aperture (*D*=30 cm). Using (3) and (4), the theoretical resolution limits of the synthesized aperture were calculated to be $\delta_{cr}$=0.94 cm and $\delta_{cr}$=1.67 cm, respectively.

Accordingly, the scene discretization voxel size for imaging of the point-scatter target was selected to be $\Delta y=\Delta z=1$ cm in cross-range and $\Delta x=1.5$ cm in range, respectively. The least-squares reconstructed image of the target is shown in Fig. 7b. The reconstructed image in Fig. 7b was up-sampled by a factor of two for plotting.

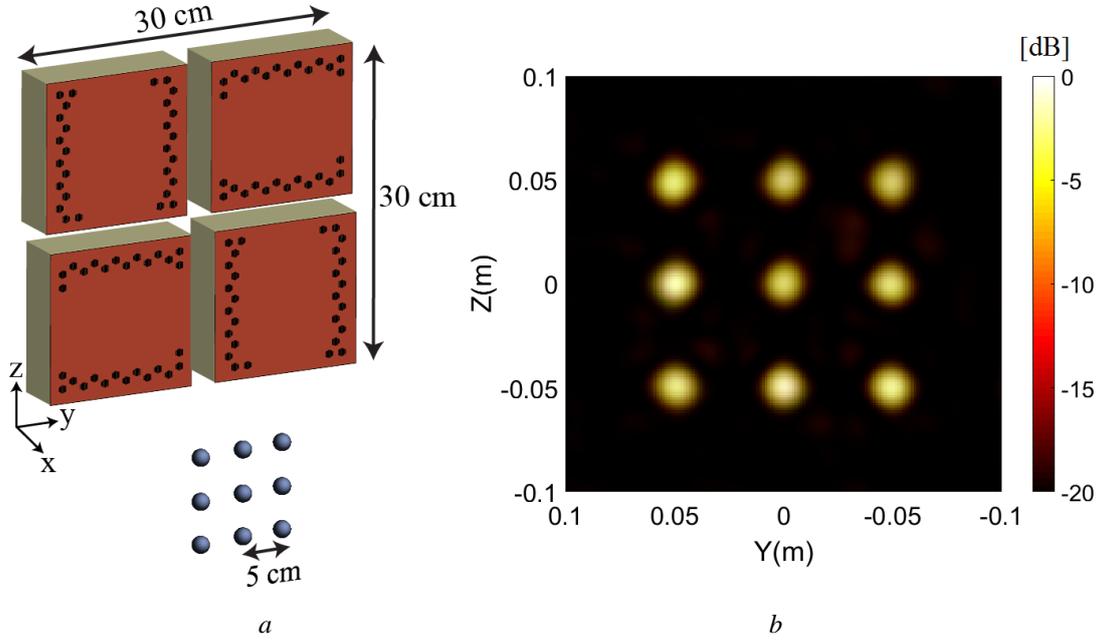

**Fig. 7.** *Imaging of a point-scatter array*
*a* Synthesized composite aperture and system layout
*b* Reconstructed image of the target

Analyzing the -3 dB full-width-half-maximum (FWHM) of the PSF pattern demonstrated in Fig. 7b, the resolution of the aperture was measured to be $\delta_{cr}=1$ cm in cross-range and $\delta_r=1.68$ cm in range, respectively. These limits exhibit good agreement with the theoretical limits calculated using (3) and (4) above, confirming that the imaging is done at the diffraction limit of the synthesized aperture.

The frequency-diverse aperture shown in Fig. 7 was synthesized using the near-field scans of the Mills-Cross antennas 3D printed using the Electrifi conductive polymer material, which has a conductivity of $\sigma=1.67*10^4$ S/m. As previously shown in Fig. 5a, the material conductivity value for 3D printing has a significant effect on the impulse response (and the Q-factor) of the antennas. To put this statement into an imaging perspective, we synthesized the same frequency-diverse aperture shown in Fig. 7a but varied the Q-factor of the antennas as a function of the conductivity of the Electrifi material for 3D printing. To this end, three frequency-diverse apertures were synthesized. In the first aperture, the frequency-diverse antennas have a Q-factor of $Q=150$, corresponding to $\sigma=1.67*10^3$ S/m. In the second aperture, the frequency-diverse antennas have a Q-factor of $Q=300$, corresponding to $\sigma=1.67*10^4$ S/m. And finally, in the third aperture, the frequency-diverse antennas have a Q-factor of $Q=800$, corresponding to $\sigma=1.67*10^5$ S/m. In each scenario, the synthesized aperture images a 1.5 cm resolution target, consisting of vertical and horizontal stripes of 1.5 cm width that are separated by 1.5 cm distance from each other (selected in accordance with the resolution limit of the aperture). The least-squares reconstructed images of the resolution target are shown in Figs. 8a-8c.

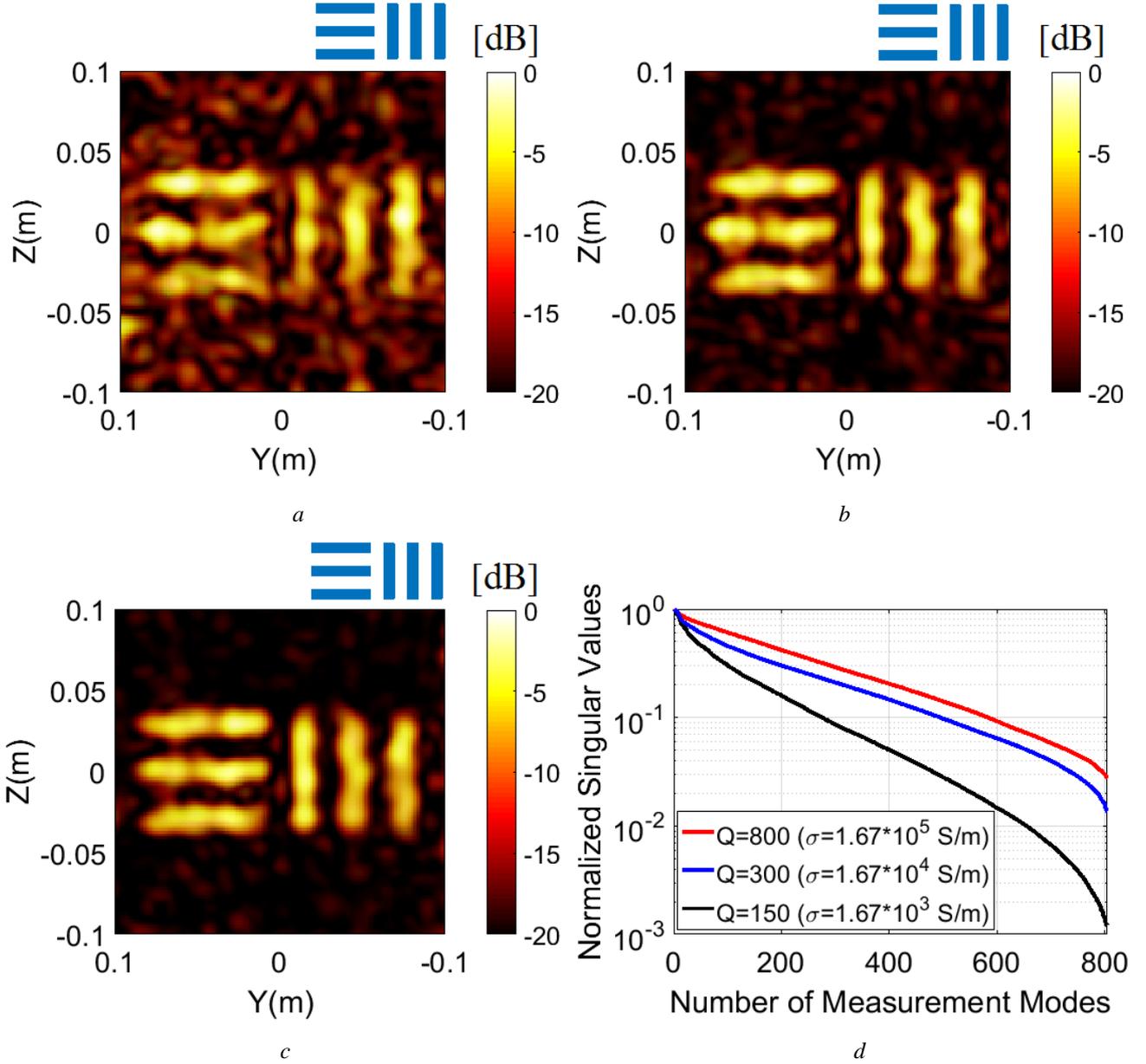

*Fig. 8. Reconstructed images of the resolution target as a function of antenna Q-factor and material conductivity. The imaged actual resolution target is shown in the top right corner of the reconstructed images*
*a* $Q=150$ ($\sigma=1.67*10^3$ S/m)
*b* $Q=300$ ($\sigma=1.67*10^4$ S/m)
*c* $Q=800$ ($\sigma=1.67*10^5$ S/m)
*d* SVD pattern of the measurement matrix, **H**

As shown in Figs. 8*a*-8*c*, increasing the conductivity of the Electrifi material (and therefore the Q-factor of the 3D printed antennas) improves the fidelity of the reconstructed image. The conditioning of the inverse problem defined in (1) can be analyzed by means of a singular value decomposition (SVD) analysis of the measurement matrix, **H** [21, 46]. In Fig. 8*d*, we demonstrate the SVD patterns of **H** as a function of material conductivity and the corresponding antenna Q-factor. It is evident in Fig. 8*d* that increasing the conductivity of the Electrifi material results in a superior conditioning of the measurement matrix, **H.** This is the underlying reason behind the improvement in the reconstructed images shown in Figs. 8*a*-8*c* as the conductivity value of the material is increased.

## 4. Conclusion

By harnessing the 3D printing technology and recent advances in material engineering, we have demonstrated the application of 3D printed frequency-diverse antennas for microwave imaging applications. The fabrication of the antennas has been achieved by means of a simple 3D printing process using a combination of PLA and conductive polymer (Electrifi) materials, circumventing the need for additional metallization and other conventional machining, photolithography and laser fabrication techniques. It has been demonstrated in the Virtualizer that using the frequency-diverse aperture synthesized with the 3D printed antennas, good quality images of objects have been achieved by means of a simple frequency sweep over the K-band. It has also been shown by full-wave simulations that the performance of the 3D printed Mills-Cross cavity antennas could be further improved by increasing the conductivity of the Electrifi polymer material. This is an ongoing research effort with the initial results suggesting that an increase in the material conductivity by a factor of 10 can be achieved. The proposed technology holds significant potential in a number of applications where custom made antenna equipment with low-cost and rapid manufacturing is required, such as security-screening, biomedical imaging, non-destructive testing, body-centric communications and wireless power transfer applications.

## Acknowledgments

This work was supported by the Air Force Office of Scientific Research (AFOSR, Grant No. FA9550-12-1-0491).